\documentclass[twocolumn,aps,pra]{revtex4}
\usepackage{epsfig}
\usepackage[english]{babel}
\usepackage{latexsym}
\usepackage{graphics}
\usepackage{subfigure}
\usepackage{graphicx}
\usepackage{dcolumn}
\usepackage{amsmath}
\usepackage{hyperref}
\usepackage{amssymb}
\usepackage{color}

\begin{document}
\title{Coulomb scattering inducing time lag in strong-field tunneling ionization}

\author{Y. G. Peng$^{1,\ddag}$, J. Y. Che$^{1,\ddag}$, C. Chen$^{1}$,  G. G. Xin$^{2,\dag}$, and Y. J. Chen$^{1,*}$}

\date{\today}

\begin{abstract}
We study ionization of atoms in strong elliptically-polarized laser fields.
We focus on the physical origin of the offset angle in the photoelectron momentum distribution and its possible relation to a specific time.
By developing a model which is based on strong-field approximation and considers the classical Coulomb scattering,
we are able to quantitatively explain recent attoclock experiments in a wide region of laser and atomic parameters.
The offset angle can be understood as arising from the scattering of the  electron by the ionic potential
when the electron exits the laser-Coulomb-formed barrier through tunneling.
The scattering time is manifested as the Coulomb-induced ionization time lag and is encoded in the offset angle.

\end{abstract}

\affiliation{1.College of Physics and Information Technology, Shaan'xi Normal University, Xi'an, China\\
2.School of Physics, Northwest University, Xi'an, China}

\maketitle

\section{Introduction}
When atoms are exposed to strong elliptically-polarized laser fields with high ellipticity,
the photoelectron momentum distributions (PMDs) show a ring-like structure.
In comparison with the theoretical predictions of the well-known strong-field approximation (SFA) \cite{Lewenstein1995}
where the Coulomb potential is neglected, the experimental PMDs  present a small rotation. The offset angle $\theta$ in PMD
characterizes this rotation. As this offset angle arises from the Coulomb effect and depends on laser and atomic parameters,
experimental techniques have been developed to use this angle to  probe attosecond tunneling dynamics of the electron in recent years.
Relevant techniques have been termed as attosecond angle streaking or attoclock \cite{Eckle1,Eckle2,Eckle3,Eckle4}. Many attoclock experiments have been performed for probing the tunneling time (i.e., the time spent by the electron under the barrier during the process of tunneling).
As some experiments show that tunneling needs a finite time \cite{Landsman,Camus,Teeny}, other show that tunneling is instantaneous  \cite{Undurti,Quan,Klaiber,Torlina}.

To obtain the time information from the measured offset angle, one needs to establish a definite mapping
between the concerned time and this angle. To do so, one first need to develop a time-dependent analytical model
which includes the Coulomb potential to describe the ionization dynamics of atoms in a strong laser field.
Then this mapping can be deduced from the developed model. Some Coulomb-included strong-field models have been established in recent years \cite{MishaY,Goreslavski}, such as the Coulomb-modified SFA (CM-SFA) \cite{yantm2010} which considers the Coulomb effect
after the tunneling electron exits the barrier with numerical solution of the Newton equation,
the analytical R-matrix (ARM) theory \cite{Torlina}, which considers the Coulomb effect
when the tunneling electron is under the barrier with including the Coulomb potential into the saddle-point equation,
the time-free Keldysh-Rutherford (KR) model \cite{Bray} which considers the Coulomb effect in terms of classical
Coulomb scattering to explain the origin of the offset angle in the case of low laser intensity, the extended KR model \cite{Serov}
which considers the Coulomb effect with the procedure somewhat similar to CM-SFA but solves the Newton equation analytically, etc..
Due to the complexity in analytical treatment of the Coulomb problem in strong-field ionization,
some approximations have to be performed in actual manipulation. These approximations can influence
predictions of models for the offset angle and the deduction of the concerned time by this angle.

Very recently, a strong-field model termed as tunneling-response-classical-motion (TRCM) model \cite{Chen2021}
has also been developed to
explain the origin of the offset angle in terms of the response time of the electron inside an atom to light in strong-laser induced
photoelectric effect. This response time can be understood as  the observable duration time of strong three-body interaction between electron,
nucleus and photon and is characterized by the Coulomb-induced ionization time lag \cite{Xie,Che2}.
Although the TRCM can quantitatively and consistently explain recent attoclock experiments with different laser and atomic parameters,
the physical picture in TRCM related to the response time is somewhat complex.

In this paper, we develop a strong-field model by including the Coulomb scattering into the general SFA (CS-SFA).
Our model is able to not only explain a series of recent attoclock experiments in a wide parameter region but also provide
an intuitive physical picture for the origin of the offset angle. We show that when the electron exits the laser-Coulomb-formed barrier
through tunneling with a certain momentum, due to the existence of the Coulomb potential, the electron is not ionized immediately.
Instead, it is elastically scattered by the Coulomb potential in a short time (the scattering time), resulting in a momentum shift
and an ionization time lag.
This momentum shift quantifies the offset angle and the scattering time explains the ionization time lag.

\section{Coulomb-scattering SFA}

\begin{figure}[t]
\begin{center}
\rotatebox{0}{\resizebox *{7.5cm}{7cm} {\includegraphics {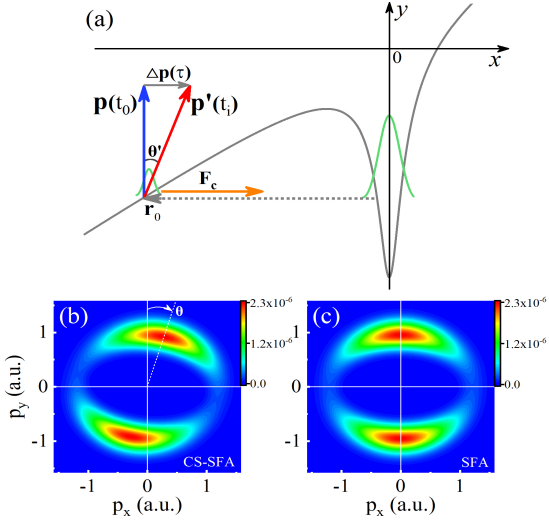}}}
\end{center}
\caption{A sketch of the Coulomb scattering picture described in CS-SFA.
When the tunneling electron exits the barrier at the laser-field peak time $t_0$ and
at the position $\textbf{r}_0$ with a momentum $\textbf{p}(t_0)$,
it is scattered by the Coulomb force $\textbf{F}_c$ for a small period of time $\tau$, resulting in a momentum shift $\triangle \textbf{p}(\tau)$.
Then it is free at the time $t_i=t_0+\tau$ with a scattered momentum $\textbf{p}'(t_i)$.
 The scattering angle $\theta'$ between $\textbf{p}$ and $\textbf{p}'$
amounts to the offset angle $\theta$ defined in the PMD  obtained with CS-SFA for the H atom (b).
In SFA with neglecting the Coulomb potential, the tunneling electron escapes
with the momentum $\textbf{p}(t_0)$ and the PMD obtained for H does not show the offset angle (c).
Laser parameters used in calculations are $I=2\times10^{14}$ W/cm$^{2}$, $\lambda=800$ nm and $\epsilon=0.84$.
}
\label{fig:g1}
\end{figure}

\emph{SFA}.-We begin our discussions with SFA where the tunneling mechanism is considered to dominate in strong-field ionization  \cite{Keldysh,Faisal,Reiss}.
In the length gauge, the amplitude of the photoelectron with the drift momentum $\textbf{p} $ can be written as \cite{Lewenstein1995}
\begin{equation}
c(\textbf{p})=-i\int^{T_{p}}_0dt^\prime{\textbf{E}}(t^\prime)\cdot{\textbf{d}_i}{[\textbf{p}+\textbf{A}(t^\prime)]}e^{iS(\textbf{p},t^\prime)}.
\end{equation}
Here, the term  $S(\textbf{p},t')=\int_{}^{t'}\{{[\textbf{p}+\textbf{A}(t''})]^2/2+I_p\}dt''$ is the semiclassical action and $T_{p}$ is the length of the total pulse.
The term $\textbf{d}_i(\textbf{v})=\langle{\textbf{v}}|\textbf{r}\vert{{0}\rangle}$ is the transition matrix element between the ground state $\vert{{0}\rangle}$ and the continuum state $\vert{{\textbf{v}}\rangle}$, and  $\textbf{A}(t)$ is the vector potential of the electric field $\textbf{E}(t)$.
The temporal integral in the expression of $c(\textbf{p})$ can be evaluated by the saddle-point method \cite{Lewenstein1995,Becker2002},
with the solution of the following equation
\begin{equation}
[\textbf{p}+\textbf{A}(t_s)]^2/2=-I_p.
\end{equation}
The solution $t_s=t_0+it_x$ of the above equation is complex and the real part $t_0$ can be understood as the
tunneling-out time at which the tunneling electron exits the barrier. Without considering the Coulomb potential,
the tunneling-out time $t_0$ also amounts to the ionization time at which the electron is free.

From a semiclassical view of point, at the tunnel exit with the time $t_0$,
the exit position $\textbf{r}_0$ of the tunneling electron can be evaluated with
$\mathbf{r}_0\equiv\mathbf{r}(t_0)=Re(\int^{t_0}_{t_0+it_{x}}[\mathbf{p}+\mathbf{A}(t')]dt')$,
and the exit velocity can be evaluated with $\textbf{v}(t_0)=\mathbf{p}+\mathbf{A}(t_0)$.
Note, according to Eq. (2), the exit velocity $\textbf{v}(t_0)$ also agrees with the relation of
$\textbf{v}(t_0)\approx-Re[\textbf{A}(t_s)]+\textbf{A}(t_0)$. It reflects the basic quantum effect of the system in tunneling.
This velocity expression also gives the mapping between the drift momentum $\mathbf{p}$ and the ionization time $t_0$ in SFA. That is
\begin{equation}
\textbf{p}\equiv\textbf{p}(t_0)=\textbf{v}(t_0)-\textbf{A}(t_0).
\end{equation}

\emph{CS-SFA}.-Next, we introduce the effect of the Coulomb potential $V(\textbf{r})$ into the SFA. At the tunnel exit with $r_0\sim 10$ a.u.
for general laser and atomic parameters used in experiments,
the electron is not far away from the nucleus and it is also subject to the non-negligible Coulomb force $\textbf{F}_c=-\frac{Z}{r^{3}_0}\textbf{r}_0$.
Here, $Z$ is the nuclear charge
and ${r}_0=|\textbf{r}_0|$. Therefore, when the Coulomb effect is included,
the electron can not be considered to be free immediately at the tunneling-out time $t_0$.

\emph{Coulomb scattering}.-With the above discussions, we assume that when the electron exits the barrier at  $t_0$ with the exit position $\textbf{r}(t_0)$ and the exit velocity $\textbf{v}(t_0)$, it is immediately scattered by the Coulomb potential.
During the scattering process  with the duration of a small period of time $\tau$, the electron momentum changes from $\textbf{p}$ to $\textbf{p}'$,
then it is free at the time $t_i=t_0+\tau$.
The scattered momentum $\textbf{p}'$ and the ionization time $t_i$ agrees with the following relation
\begin{equation}
\textbf{p}'\equiv\textbf{p}'(t_i)=\textbf{v}(t_0)-\textbf{A}(t_i).
\end{equation}
The above expression implies that due to the existence of the Coulomb potential, the electron is born at the time $t_i$
with a Coulomb-induced time lag  $\tau$ to the Coulomb-free birth time $t_0$  and with the tunneling-induced exit velocity $\textbf{v}(t_i)=\textbf{v}(t_0)$.
The momentum shift $|\triangle \textbf{p}|=|\textbf{p}'-\textbf{p}|$ can be determined by the classical Coulomb scattering formula with the assumption of a elastic scattering from $\textbf{p}$ to $\textbf{p}'$. That is  \cite{Bray,Landau}
\begin{equation}
|\triangle\mathbf{p}|=2|\mathbf{p}|\sin(\frac{\theta'}{2})=\frac{2Z}{pr_b}
\end{equation}
with $p=|\textbf{p}|\approx|\textbf{p}'|$, $\sin(\frac{\theta'}{2})=Z/({{p}}^2r_b)$ and $r_b\approx r_0$.
Here, $\theta'$ is the scattering angle.
The elastic-scattering assumption is applicable for the condition of $|V(r_0)|=Z/{r_0}\ll  {{{\mathbf{p}}^{2}}}/{2}$.
By Eq. (3) and Eq. (4), we also have
\begin{equation}
|\triangle\mathbf{p}|\equiv|\triangle\mathbf{p}(\tau)|=|\textbf{A}(t_i)-\textbf{A}(t_0)|.
\end{equation}
Therefore, as the momentum shift  $|\triangle \textbf{p}|$ is determined by Eq. (5), we can also obtain the time lag $\tau=t_i-t_0$.

\emph{Analyses}.-The offset angle $\theta$ is related to
the electron which exits the barrier at the peak time of the laser field 
and therefore has the maximal amplitude.
According to our Coulomb-scattering presumption,
in this case, the scattering angle $\theta'$ of the electron equals to the offset angle $\theta$.
We assume that the laser field $\textbf{E}(t)$ has the form of $\textbf{E}(t)=\textbf{e}_x{E}_x(t)+\textbf{e}_y{E}_y(t)$, where $E_{x}(t)=E_0\sin(\omega t)$, $E_{y}(t)=E_1\cos(\omega t)$,
$E_0={E_L}/{\sqrt{1+\epsilon^2}}$ and $E_1=\epsilon E_0$, with $E_L$ being the maximal laser amplitude related to the peak intensity $I$, $\epsilon$  the ellipticity, $\omega$  the laser frequency.
Without loss of generality, we limit our discussions to the first half laser cycle where the offset angle is related to the time $t_0$ agreeing with $\omega t_0=\pi/2$. At this time,  Eq. (2) tells that $|\textbf{p}|=p_y$ and the vector $\textbf{r}_0$
(and therefore the Coulomb force $\textbf{F}_c$) is along the $x$ axis.
When the scattering angle $\theta'$ is small, we also have
$|\triangle\mathbf{p}|\approx  p'_x$, ${p}'_y\approx  p_y$ and
$\tan(\frac{\theta}{2})=\tan(\frac{\theta'}{2})\approx\sin(\frac{\theta'}{2})\approx Z/({{p}}^2r_0)$.
A sketch of the scattering picture described in the CS-SFA is presented in Fig. 1.

\emph{Offset angle formula}.-According to the above discussions, the offset angle $\theta$ can be written as
\begin{equation}
\theta\approx\arctan(\frac{2Z}{{{{p}}^{2}}r_0})
\end{equation}
Here, $p=|\textbf{p}|=p_y={v_y}(t_0)-{A_y}(t_0)$. As the momentum $p$ considers the nonzero exit velocity $v_y(t_0)$, we call
the expression of $\theta$ the nonadiabatic one. Note, for the peak time $t_0$ of $E_x(t)$ related to the offset angle, $|v_x(t_0)|=0$.
Accordingly, considering that $\tau$ is a small quantity, at the peak time $t_0$, we also have $|\textbf{A}(t_i)-\textbf{A}(t_0)|\approx E_0\tau$.
Then according to Eq. (5) and Eq. (6), we  also have
\begin{equation}
\tau\approx\frac{2Z}{E_0{{{p}}}r_0}
\end{equation}
with $p=p_y={v_y}(t_0)-{A_y}(t_0)$. Considering that for hydrogen-like atoms
$Z\approx\sqrt{2I_p}$ and assuming that
$r_0\approx I_p/E_0$  as in \cite{Bray}, we also have
\begin{equation}
\theta\approx\arctan(\frac{2\sqrt{2}E_0}{{{{p^{2}_y}}}\sqrt{I_p}})
\end{equation}
with $\tau\approx{2\sqrt{2}}/({{{{p_y}}}\sqrt{I_p}})$.
In the paper, we will perform nonadiabatic calculations using the above expression.

\emph{Adiabatic approximation}.-By neglecting the term $v_y(t_0)$ in the expression of $p_y$,
we can also obtain the adiabatic description of $\theta$ and $\tau$.
With the further consideration of $\tan\theta\approx\theta$ for a small angle $\theta$ and
$p_y\approx-{A_y}(t_0)=\epsilon E_0/\omega$,
we have
\begin{equation}
\theta\approx\frac{2\sqrt{2}\omega^2}{{\epsilon^2E_0}\sqrt{I_p}}
\end{equation}
with $\tau\approx{2\sqrt{2}\omega}/({{{{\epsilon E_0}}}\sqrt{I_p}})$. From Eq. (10), we also see $\theta\approx\omega\tau/\epsilon$.
This is just the empirical relation between the angle $\theta$ and a time $\tau$ given in attoclock.
Here, this relation is obtained from the adiabatic version of CS-SFA with giving clear physical definition of $\theta$ and  $\tau$,
namely the Coulomb-scattering angle $\theta$ and the scattering time $\tau$.
In the paper, we will perform relevant adiabatic calculations  using Eq. (10).

\emph{PMDs}.-The momentum shift $\triangle\mathbf{p}$ induced by $\textbf{F}_c$ is always contrary to the vector $\textbf{r}_0$,
as shown in Fig. 1.
Then with considering  Eq. (5), we have $\triangle\mathbf{p}=-|\triangle\mathbf{p}|\textbf{r}_0/r_0=-{2Z}\textbf{r}_0/({pr^2_0})$ and $\textbf{p}'=\textbf{p}+\triangle\mathbf{p}$. By assuming that the scattering influences only the saddle-point momentum $\textbf{p}\equiv \textbf{p}(t_0)$ of Eq. (3) for arbitrary time $t_0$ and does not
change the corresponding amplitude  $c(\textbf{p})$, we can obtain the CS-SFA amplitude  $c(\textbf{p}')$ directly from the SFA one
with $c(\textbf{p}')\equiv c(\textbf{p})$ at $r_0\approx I_p/|\textbf{E}(t_0)|$.
This CS-SFA therefore provides a simple approach for studying the Coulomb effect in strong-field ionization of atoms.
The CS-SFA prediction of the PMD for H is presented in Fig. 1(b).

\begin{figure}[t]
\begin{center}
\rotatebox{0}{\resizebox *{7.5cm}{6cm} {\includegraphics {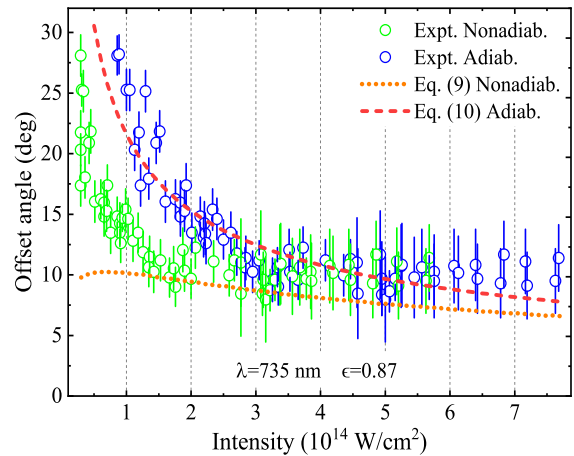}}}
\end{center}
\caption{Application to He for predicting the offset angle. Dots: experimental results with the nonadiabatic (green circle) or
 adiabatic (blue circle) laser-intensity calibration in \cite{Boge}.
Lines: predictions of Eq. (9) of nonadiabatic CS-SFA (orange dotted) and Eq. (10) of adiabatic one (red dashed).
Laser parameters used are as shown.}
\label{fig:g2}
\end{figure}

\section{Results and discussions}

\emph{Applications to different targets}.-We first show the predictions of CS-SFA for the offset angle of the He atom with comparing to experiments \cite{Boge}.
Due to the uncertain in calibrating the laser intensity in experiments, the  experimental results are presented with nonadiabatic and adiabatic
calibrating procedures, corresponding to Eq. (9) and Eq. (10) in our treatments, respectively.
Firstly, for nonadiabatic cases, the predictions of Eq. (9) agree with the experimental results for higher laser intensities with
$I>1.2\times10^{14}$ W/cm$^{2}$.
For lower laser intensities,
the experimental angles increase very fast with the decrease of the intensity, while the model ones change slowly.
When the laser intensity is low, the quantum effect beyond tunneling can play an important role  in ionization.
In this case,
the present tunneling-based treatment may not give a full description for the Coulomb effect.
For adiabatic cases, the results of Eq. (10) agree with the experimental data on the whole.
Since  Eq. (10) is an approximation to  Eq. (9) in CS-SFA.
In the following discussions, we focus on the results of Eq. (9).

\begin{figure}[t]
\begin{center}
\rotatebox{0}{\resizebox *{8cm}{6cm} {\includegraphics {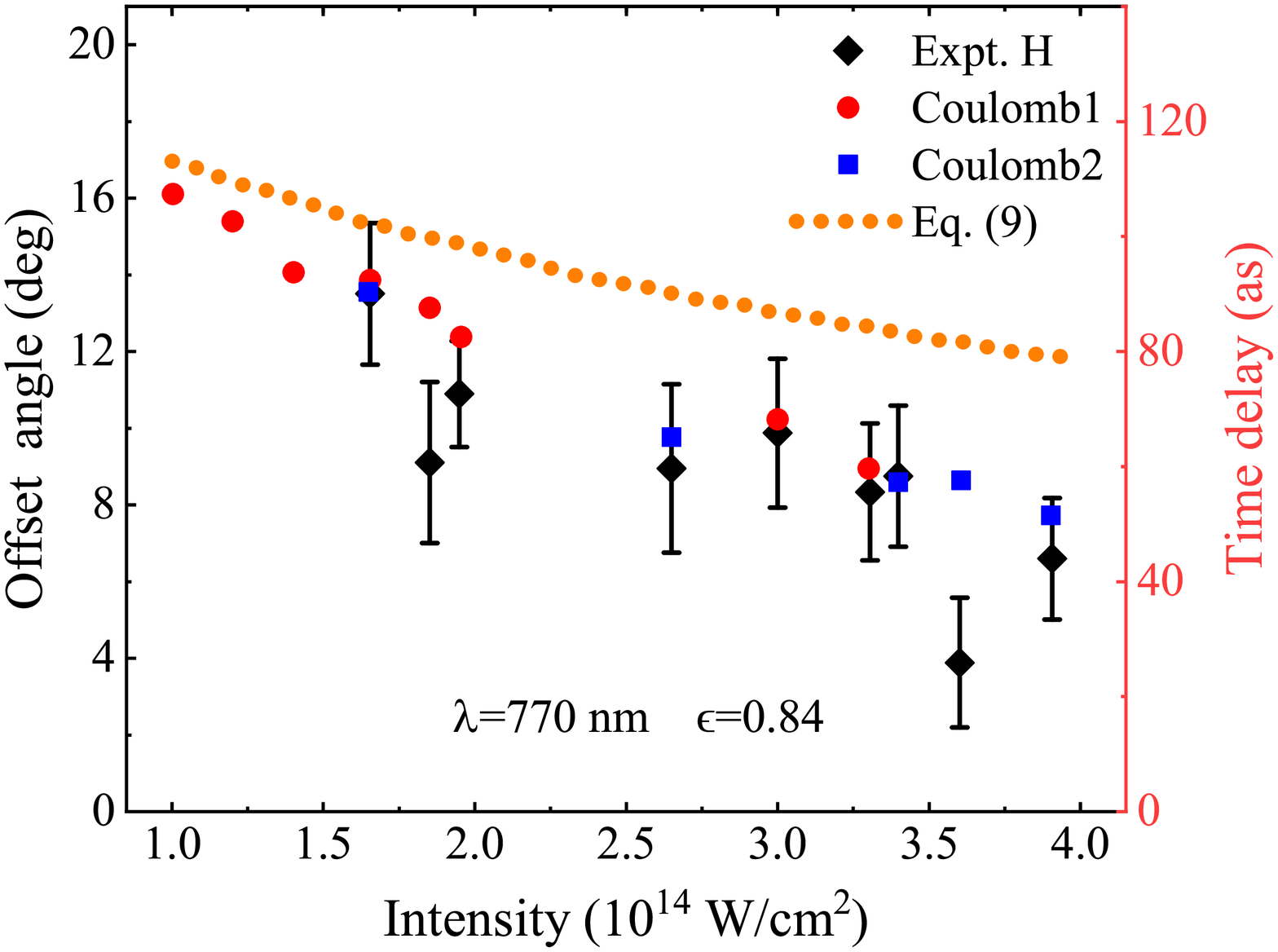}}}
\end{center}
\caption{Application to H for predicting the offset angle and the time delay.
Dots: experimental (black diamond) and 3D-TDSE (red circle and blue square) results  in \cite{Undurti}.
Orange line: predictions of Eq. (9) of nonadiabatic CS-SFA. Laser parameters used are as shown.
The time delay $\tau$ (right axis) is obtained from the offset angle $\theta$ with $\theta\approx\omega\tau/\epsilon$.
}
\label{fig:g3}
\end{figure}

In Fig. 3, we apply our model to the H atom, with comparing to experimental and 3D-TDSE results in  \cite{Undurti}.
For lower laser intensities with $I<2\times10^{14}$ W/cm$^{2}$, the predictions of Eq. (9) are about 1 degree to 2 degree larger than the experimental and TDSE results. The difference between them becomes somewhat larger at higher laser intensities.
For high intensities, the ionization of H with  $I_p=0.5$ a.u. is strong and therefore the ground-state depletion is also remarkable.
This remarkable depletion is not considered in the present SFA-based scattering model.
We mention that in \cite{Undurti}, the time delay $\tau$ is
obtained from the angle $\theta$ with the relation $\theta\approx\omega\tau/\epsilon$.
This relation holds in our theory in the adiabatic approximation. Here, for direct comparisons to results in  \cite{Undurti},
we also use the relation to evaluate the delay $\tau$ from the angle $\theta$ of Eq. (9).

We further apply our model to more targets such as He and Ar \cite{Eckle3}, H$_2$ \cite{Quan} and H \cite{Torlina}
at a wider range of laser parameters.
Relevant results are presented in Fig. 4 with comparing to real and numerical experiments.
Equation (9) gives a good description for the experimental results of He and Ar in Figs. 4(a) and 4(b).
It is also able to give a good explanation for the experimental results of a small molecule H$_2$ in Fig. 4(c).
The predictions of Eq. (9) also agree with the TDSE results for H obtained in other numerical experiments
from $I=0.5\times10^{14}$ W/cm$^{2}$ to $I=2.5\times10^{14}$ W/cm$^{2}$, as seen in Fig. 4(d).
Here, for cases of lower and higher laser intensities,
the model results differ remarkably from the TDSE ones, similar to those cases discussed in Fig. 2 and Fig. 3.

\begin{figure}[t]
\begin{center}
\rotatebox{0}{\resizebox *{8.5cm}{6cm} {\includegraphics {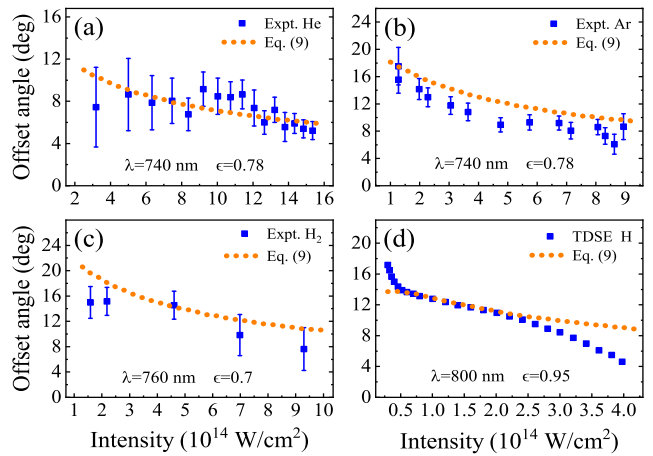}}}
\end{center}
\caption{Application to more cases for predicting the offset angle.
Blue square dots: experimental  results for He (a) and Ar (b) in \cite{Eckle3}
and for H$_2$ (c) in \cite{Quan}, and 3D-TDSE results  for H (d) in \cite{Torlina}.
Orange lines: predictions of Eq. (9) of nonadiabatic CS-SFA.
Laser parameters used are as shown.
}
\label{fig:g4}
\end{figure}

\emph{Comparisons to other theories}.-In comparison with CS-SFA  \cite{yantm2010}, the extended KR model \cite{Bray}
and the ARM theory \cite{Torlina},
the CS-SFA considers the Coulomb effect with classical Coulomb scattering.
In comparison with the time-free KR model \cite{Bray},
the CS-SFA is time dependent and considers
the nonzero exit velocity related to quantum effects in tunneling.
In comparison with TRCM approach \cite{Chen2021}, both CS-SFA and TRCM consider the Coulomb effect with an emphasis
on the central symmetry of the atomic Coulomb potential near the nucleus. 
The tunnel exit $r_0\approx I_p/E_0$ is about 10 a.u. in general cases. Around this distance,
the wave function of bound eigenstate of the atomic system with higher energy has large amplitudes.
The TRCM  therefore assumes that at the tunnel exit,
the tunneling electron is still located at a bound state (or a bound wave packet consisted of high-energy bound states)
which approximately agrees with the virial theorem. A small period of time $\tau$ (the response time) is needed for the electron to
move from the bound state to a continuum state. Instead of introducing the bound state, the CS-SFA assumes a
Coulomb-induced classical elastic scattering at the tunnel exit. In this case, the electron also spends
a small period of time $\tau$ (the scattering time)
to move from a Coulomb-free emitting state to a scattering state.
For most of cases discussed here, the predictions of TRCM and CS-SFA are comparable,
suggesting that the somewhat abstract response time can be understood with the intuitive scattering time.

\section{conclusion}

In conclusion, with introducing the classical Coulomb scattering into the SFA at the tunnel exit,
we have constructed a simple model to consider the Coulomb effect in strong-field ionization.
In our model, the offset angle can be attributed to the scattering angle and
the Coulomb-induced ionization time lag can be understood as the scattering time,
providing a clear time-resolved physical picture for the origin of the offset angle.
Simple formula have  been derived for the offset angle and the ionization time lag,
and a definite mapping between this angle and this time has also been established.
Our model is able to quantitatively explain a series of recent attoclock experiments for intermediate laser intensities.
It may be helpful for understanding and studying experimental phenomena in attosecond-resolved  measurements.

This work was supported by the National Natural Science Foundation of China (Grant No. 12174239),
and the Fundamental Research Funds for the Central Universities of China (Grant No. 2021TS089).

\end{document}